\RequirePackage{fix-cm}
\documentclass[a4paper,conference]{IEEEtran}
\usepackage[latin9]{inputenc}
\usepackage{amsmath}\usepackage{amssymb}\usepackage{graphicx}
\usepackage{tikz}\usetikzlibrary{automata,arrows,positioning,calc}

\makeatletter

\usepackage{epsfig}\usepackage{amsfonts}\usepackage{array}\usepackage{dcolumn}\usepackage{psfrag}\usepackage{amsfonts}\usepackage{accents}\usepackage{booktabs}
\usepackage{cite}\usepackage{subfigure}\usepackage[all]{xy}
\usepackage{dsfont}\usepackage{url}\usepackage{xcolor}\usepackage[nolist]{acronym}
\usepackage[footnote,draft,silent,nomargin]{fixme}
\usepackage{algorithm}
\usepackage[noend]{algpseudocode}
\usepackage{amsthm}
\usepackage{mathtools}
\usepackage{xfrac}
\usepackage{booktabs}
\usepackage{longtable}
\usepackage{balance}

\graphicspath{{./}}

\acrodef{SINR}{Signal to Interference and Noise Ratio}
\acrodef{CDF}{Cumulative Distribution Function}
\acrodef{PDF}{Probability Density Function}
\acrodef{MTC}{Machine-type Communications}
\acrodef{M2M}{Machine-to-Machine}
\acrodef{DTMC}{Discrete-Time Markov Chain}
\acrodef{RV}{Random Variable}
\acrodef{MTTR}{Mean Time to Restoration}
\acrodef{TTI}{Transmission Time Interval}

\definecolor{BuGn}{RGB}{28,144,153}

\makeatother
\begin{document}

\title{Reliability and Error Burst Length Analysis of Wireless Multi-Connectivity}

\author{
Jimmy~J.~Nielsen, Israel Leyva-Mayorga and Petar~Popovski\\
Connectivity section, Department of Electronic Systems, Aalborg University, Denmark\\
\{jjn,ilm,petarp\}@es.aau.dk\\%
}

\maketitle
\begin{abstract} 
Multi-connectivity offers diversity in terms of multiple interfaces through which the data can be sent, thereby improving simultaneously the overall reliability and latency. This makes interface diversity a natural candidate for supporting Ultra-Reliable Low Latency Communications (URLLC). This work investigates how the packet error statistics from different interfaces impacts the overall reliability-latency characteristics. We use the simple Gilbert-Elliott model for burst errors and estimate its parameters based on  experimental measurement traces from LTE and \mbox{Wi-Fi} packet transmissions collected over several days. The results show that using interface diversity configurations that include at least one \mbox{Wi-Fi} interface leads to, somewhat surprisingly, since Wi-Fi is generally less reliable than LTE, superior results in terms of packet success and error burst duration. Another interesting finding is that \mbox{Wi-Fi}-based interface diversity configurations outperform even ultra-reliable single links.
\end{abstract}

\section{Introduction}
One of the main connectivity types for the fifth generation of mobile networks (5G) is Ultra-Reliable and Low-Latency Communication (URLLC). Reliability and latency requirements for this use case are in the order of $1-10^{-5}$ and of a few milliseconds, respectively. An additional challenge in URLLC is that the requirement for low packet error probability is coupled with the stringent latency requirements. For instance, current cellular systems incorporate mechanisms, such as the hybrid automatic repeat request (HARQ), that provide an extremely high degree of reliability, but that cannot guarantee stringent latency requirements. 

Today a typical smartphone incorporates numerous wireless interfaces that can be used to establish an equal number of communication paths; this is commonly known as multi-connectivity~\cite{Wolf2019}. \emph{Interface diversity}~\cite{Nielsen2018} is a specific way of utilizing the multi-connectivity in order to increase the reliability of the transmission by adding redundancy. This approach is particularly appealing to support URLLC or URLLC-like connectivity, as it can be employed to simultaneously increase the end-to-end reliability and to reduce the latency.

In previous work~\cite{Nielsen2018} we studied the benefits of interface diversity in terms of reliability for a given error probability. However, errors in the wireless links oftentimes occur in bursts. This behavior highly impacts URLLC applications, where the number of consecutive errors could be more important than the error probability in the stationary regime. It is in these cases where interface diversity can provide great benefits, especially when the errors between the two interfaces are lightly, if at all, correlated. 

In this paper, we study the impact of interface diversity on the burst error distribution with a receiver-transmitter pair. For this, we consider two technologies: LTE and \mbox{Wi-Fi}, and compare the burst error and success lengths with interface diversity vs. single interface communication in two scenarios. In the first one, the same latency deadline is set for all the interfaces so that all of these provide the same reliability and, in the second one, a longer latency deadline is allowed when using a single interface, thus, providing a higher reliability.

The rest of the paper is organized as follows. We present a revision of related works and the specific characteristics of interface diversity in Section~\ref{sec:IFDforURC}. Next, we present the system model in Section~\ref{sec:BEM} and the data collection experiment in Section~\ref{sec:testbed}. Then, we present the numerical results on the length of error and success bursts in Section~\ref{sec:results} and conclude the paper in Section~\ref{sec:conclusion}.

\section{Interface Diversity for ultra-reliable communications}
\label{sec:IFDforURC}
Multi-connectivity has been studied from different perspectives to enable URLLC in 5G. For instance, Wolf \emph{et al.}~\cite{Wolf2019} studied a scenario with one user equipment (UE) connected to multiple base stations (BS) and with multiple simultaneous connections to the same BS. The benefits of this approach are assessed in terms of transmit power reduction, achieved by increasing the signal-to-noise ratio (SNR). Following a similar multi-connectivity approach, a matching problem is formulated by Simsek \emph{et al.}~\cite{Simsek2019}, where the number of UEs in the network and the limited wireless resources are considered. The objective is to provide the desired reliability to numerous users by assigning only the necessary amount of resources to each of them. Mahmood~\emph{et al.}~\cite{Mahmood2018} investigated a similar problem in a heterogeneous network scenario with a small cell and a macro cell. Their results show that multi-connectivity is particularly useful for cell-edge UEs connected to the small cell, and provides even greater benefits when URLLC and enhanced mobile broadband (eMBB) traffic coexist.

In the studies mentioned above, only stationary error probabilities are considered. However, the use of different interfaces provides unique benefits for URLLC, especially when the bursty nature of wireless errors is considered. For instance, different interfaces are likely to present different burst error distributions, and the correlation of errors between different interfaces is expected to be much lower compared to the correlation between multiple links using the same wireless interface. Despite these evident benefits, little research has been conducted on interface diversity with error bursts.

Our previous work considered the use of different transmission strategies to exploit the benefits of interface diversity for URLLC~\cite{Nielsen2018}. As shown in~\cite{Nielsen2018}, in the context of URLLC it is beneficial to use the  
latency-reliability function, which stands for the probability of being able to transmit a data packet from a source to a destination with a given latency deadline. To illustrate this concept, let $L$ be the RV that defines the packet latency. Then, for a given interface $i$ and latency deadline $l$, the latency-reliability function is defined as
\begin{equation}
    F_i\left(l\right) = \Pr\left[L\leq l\mid i\right].
\end{equation}
From there, we define the probability of error for interface $i$ as
\begin{equation}
    P_e^{(i)}=1- F_i\left(l\right).
\end{equation}
It should be noted that the traditional definition of the probability of error is obtained for the case $l\rightarrow \infty$. HARQ and similar retransmission mechanisms may provide ultra-reliable communication in the latter case, but have little to no applicability when $l$ is in the order of a few milliseconds.

When we can be justifiable assume that the communication interfaces are independent and we use packet cloning where a full packet is transmitted via each of the $N$ interfaces, the end-to-end error probability can be calculated using systems reliability theory \cite{billinton1992reliability,Nielsen2018}:
\begin{equation}
    P_e^\text{E2E} = \prod_{i=1}^N (1 - F_i\left(l\right)) = \prod_{i=1}^N P_e^{(i)}.
\end{equation}

We now extend our previous work by incorporating a burst error model, described in the following section, at each interface.

\section{Burst error model}
\label{sec:BEM}
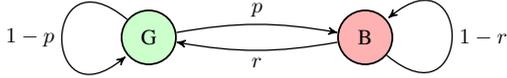
\begin{figure}[t]
	\centering
	\resizebox{0.8\linewidth}{!}{
\begin{tikzpicture}[->, >=stealth', auto, semithick, node distance=3.5cm]
\tikzstyle{every state}=[fill=white,draw=black,thick,text=black,scale=1,align=center]
\node[state]    (A)[fill=green!20]   {G};
\node[state]    (B)[right of=A,fill=red!30]   {B};
\path
 (A) edge[bend left=10]     node{$p$}         (B)
 (A) edge[in=220,out=140,loop]	    node[left]{$1-p$}         (A)
 (B) edge[bend left=10]     node{$r$}         (A)
 (B) edge[in=40,out=-40,loop]	    node[right]{$1-r$}         (B)
 
;
\end{tikzpicture}
	}                         
	\caption{Traditional two-state Gilbert-Elliott model.}
	\label{fig:2-state_GE-model}
\end{figure}

We consider the simple Gilbert-Elliott burst error model \cite{hasslinger2008gilbert} to represent a wireless interface. The model is a \ac{DTMC} that has two states: \emph{Good - G} and \emph{Bad - B}, as depicted in Fig. \ref{fig:2-state_GE-model}.
When in the good state G, packets are successfully delivered within the deadline, whereas in the bad state B, packet transmissions fail. The simple Gilbert-Elliott error model has two parameters, namely $p$ and $r$ that are used to control the resulting error probability and burst lengths~\cite{hasslinger2008gilbert}. Parameter $p$ represents a transition from state G to B and $r$ from B to G (i.e., a recovery from the bad state).

The corresponding transition probability matrix of the Gilbert-Elliott model in Fig.~\ref{fig:2-state_GE-model} is:
\begin{equation}
P = 
 	\begin{bmatrix}
    1-p & p \\
    r & 1-r
  \end{bmatrix}.
  \end{equation}

The steady-state probabilities of the good and bad states are given as \cite{hasslinger2008gilbert}:
\begin{align}
	\pi_\text{G} = \frac{r}{p+r} \\
	\pi_\text{B} = \frac{p}{r+p} \label{eq:PiB}
\end{align}
where $\pi_\text{G} + \pi_\text{B} = 1$.

Since $\pi_\text{B}$ corresponds to the fraction of time where packet transmissions fail, we can use the above relation to find matching values of $p$ and $r$ for a given packet error probability. Rewriting eq. \eqref{eq:PiB} we find:
\begin{align}
	\pi_\text{B} p + \pi_\text{B} r = p\\
	r = \frac{p}{\pi_\text{B}}-p.
\end{align}

Otherwise, the parameters can be estimated from a measurement trace. Numerous methods exist \cite{hasslinger2008gilbert}, however in this work we use the approach of Yajnik \emph{et al.}~\cite{yajnik1999measurement}. In the latter, probe packets are transmitted and their sequence numbers are recorded at the receiver. Missing sequence numbers at the receiver represent loss packets; their state is labeled as $1$, whereas the state of received packets is $0$. Let $n_0$ and $n_1$ be the number of successes and failures, respectively. Also let $n_{0 \rightarrow 1}$ be the number of times a success is followed by a failure and vice versa for $n_{1 \rightarrow 0}$. Then, the parameters are given as:
\begin{align}
	\hat{p} = n_{0 \rightarrow 1}/n_0 \nonumber\\
	\hat{r} = n_{1 \rightarrow 0}/n_1. \label{eq:est_p_r}
\end{align}

The parameters that were used for numerical evaluation were obtained from the traces of a testbed. The process for collecting these traces is described in the following.

\begin{figure*}[t]
	\centering
	\includegraphics[width=\linewidth]{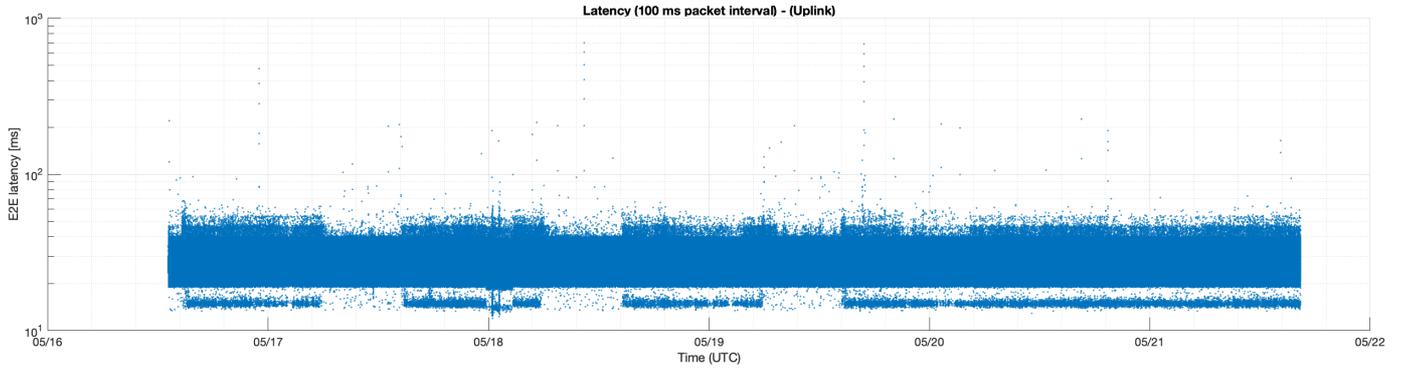}
	\caption{Multi-day trace of LTE latency measurements.}
	\label{fig:lte_trace_latency}
\end{figure*}

\begin{figure*}[t]
	\centering
	\includegraphics[width=\linewidth]{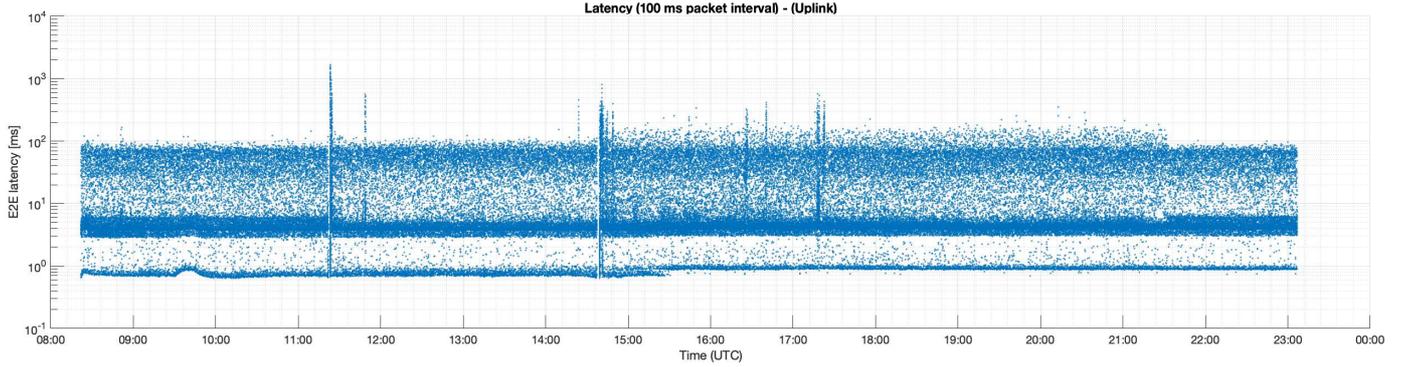}
	\caption{Single day trace of \mbox{Wi-Fi} latency measurements.}
	\label{fig:wifi_trace_latency}
\end{figure*}

\section{Latency measurements}
\label{sec:testbed}
\begin{figure}[htb]
	\centering
	\includegraphics[width=\linewidth]{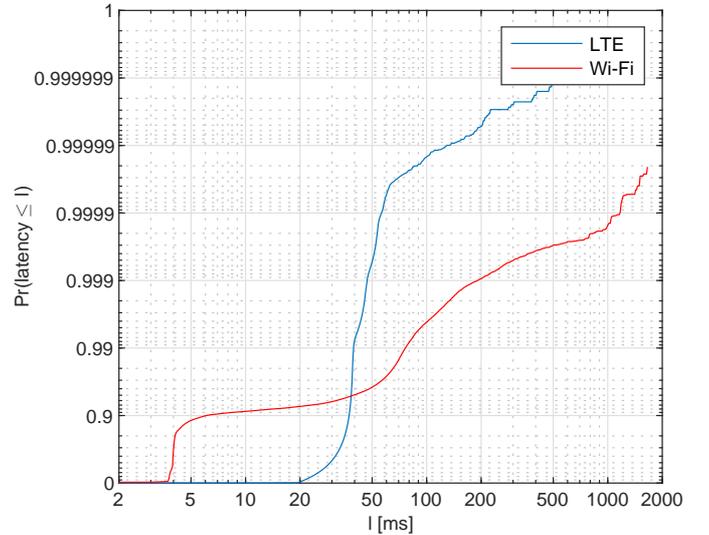}
	\caption{Empirical latency CDFs of considered interfaces.}
	\label{fig:empirical_lcdf}
\end{figure}

This section describes the data collection experiment, whose results were used to estimate the parameters for the Gilbert-Elliott model described in the previous section.
Traces of latency measurements for different communication technologies were obtained by sending small (128~bytes) UDP packets every 100~ms between a pair of GPS time-synchronized devices through the considered interface (LTE, or \mbox{Wi-Fi}) during the course of a few work days at Aalborg University campus. 
Even though the connection-less UDP transport protocol was used in the measurement campaign, the measurement traces did not reveal any actual packet losses. Any losses incurred on the inherently unreliable wireless links have been mitigated through data-link/MAC layer HARQ procedures. On the other hand, a large range in latency values was observed, as shown in the time-series plots in Fig.~\ref{fig:lte_trace_latency} and Fig.~\ref{fig:wifi_trace_latency}. In those, instances of burst errors can be identified visually as vertically aligned dots in the plots. A statistical perspective of this data is given by the latency CDFs in Fig.~\ref{fig:empirical_lcdf}, which clearly outlines some key differences between the performance of the LTE and \mbox{Wi-Fi} interfaces. While \mbox{Wi-Fi} can achieve down to 5~ms one-way uplink latency for 90\% of packets, it needs approx. 80 ms to guarantee delivery of 99\% of packets. For LTE, on the other hand, there is hardly any difference between the latency of 90\% and 99\% delivery rates, approx. 36~ms and 40~ms, respectively.
Since the measurements for both LTE and \mbox{Wi-Fi} were recorded in high-SNR radio conditions, we expect that the differences between LTE and \mbox{Wi-Fi} can, to a large extent, be attributed to the inherent differences in the protocol operation and the fact that LTE operates in licensed spectrum whereas \mbox{Wi-Fi} has to contend for spectrum access in the unlicensed spectrum.

\section{Numerical results}
\label{sec:results}
This section presents relevant results on the impact of interface diversity on the system burst error lengths with LTE and \mbox{Wi-Fi} technologies. Two cases are considered. In the first one, the latency deadline $l$ is set in such a way to achieve a similar reliability with each interface and, in the second one, the latency deadline is set differently at each of them to achieve a predefined reliability.

For analyzing the good and bad burst length distributions, we have: i) estimated the Gilbert-Elliott model parameters $p$ and $r$ for the LTE and \mbox{Wi-Fi} traces using the approach in eq. \eqref{eq:est_p_r} and ii) performed a large number of Monte Carlo simulations of the models in Matlab to acquire the burst length distributions. Each simulation results in a stochastic realization of the Gilbert-Elliott model, given the $p$ and $r$ parameters, constituted by a binary time-series $\{X(t)\}$. The outcome $X(t)=1$ corresponds to an error and $X(t)=0$ to no error in the packet transmission at time step $t$. Then, the statistics of interest can be computed and plotted.
The considered interface diversity configurations are \emph{2x LTE}, \emph{2x \mbox{Wi-Fi}}, and \emph{LTE + \mbox{Wi-Fi}}. For each of these, two independent realizations of the Gilbert-Elliott model were created as $X_1$ for interface 1 and $X_2$ for interface 2, and in each time step $t$, the diversity realization was calculated as:
\begin{equation}
	X_D(t) = \min(X_1(t),X_2(t)),
\end{equation}
meaning that the diversity configuration only fails when both of the two interfaces fail.

\subsection{Same reliability for all interfaces} 
\label{sec:burst_length_analysis}

For these experiments, the latency deadline is selected from the obtained latency CDFs, such that identical reliability is obtained at both interfaces. Hence, we select $l$ s.t.
\begin{equation}
    F_\text{LTE}(l)=F_\text{\mbox{Wi-Fi}}(l). 
\end{equation}
This is the point where latency CDF curves shown in Fig.~\ref{fig:empirical_lcdf} are crossing, i.e. approx. 95\% outage for $l=38.25$~ms. Packet delays exceeding this limit are assumed to be lost. 

\begin{table}[bt]
	\centering
	\caption{Estimated Gilbert-Elliott model parameters given $l=38.25~\textrm{ms}$.}
	\begin{tabular}{lccc}
	\toprule
		 			 	& $\hat{p}$ & $\hat{r}$ \\ \cmidrule{2-3}
		LTE   & 0.0178516 & 0.257756  \\
		\mbox{Wi-Fi} & 0.0515509 & 0.946863  \\ \bottomrule
	\end{tabular}
	\label{tab:est_p_r}
\end{table}

\begin{figure}
    \centering
    \subfigure[Good state]{\includegraphics[width=\linewidth]{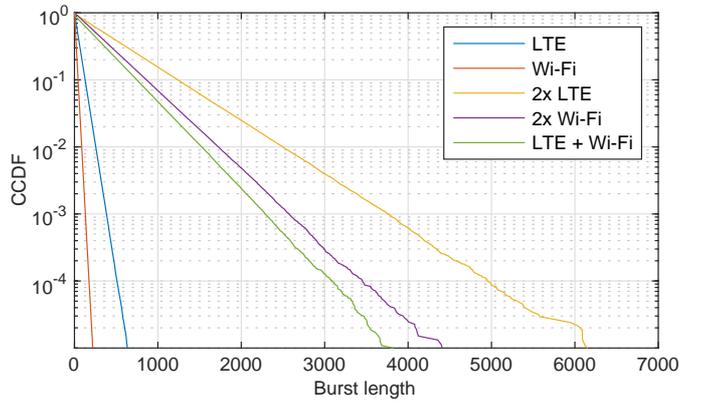}}
    \subfigure[Bad state]{\includegraphics[width=\linewidth]{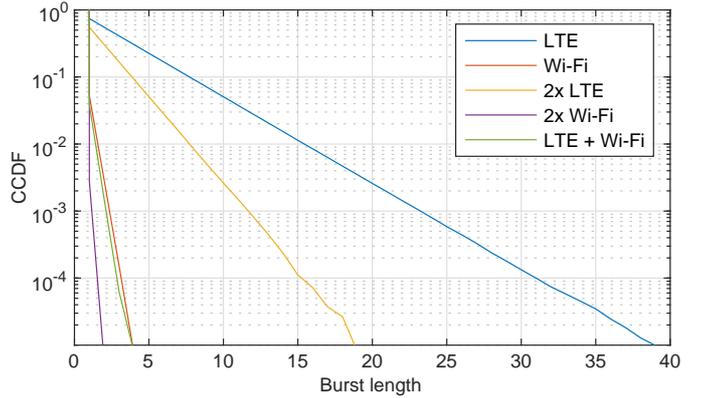}}
    \caption{Burst length distribution for Gilbert-Elliott model simulation using parameters in Table \ref{tab:est_p_r}.}
    \label{fig:burst_length}
\end{figure}

The results of the Gilbert-Elliott model analysis are shown in Fig.~\ref{fig:burst_length}. Considering initially the individual interfaces (LTE and \mbox{Wi-Fi}) in Fig.~\ref{fig:burst_length}(a), we see that LTE generally has much longer \emph{Good} burst lengths than \mbox{Wi-Fi}. Long Good burst lengths are indeed desirable, as it means that fewer interruptions occur for a given service. More critical is however the length of \emph{Bad} bursts (number of consecutive packet losses), as shown in Fig.~\ref{fig:burst_length}(b). Having bad state bursts that are ten times longer than \mbox{Wi-Fi}, LTE is inferior in this regard. Since both interfaces achieve the same packet delivery rate of approx. 95\% for $l=38.25$~ms, the suitability of each interface type depends on the service in use.

If a service is interrupted and unable to work in case of even a single or few packet losses (i.e., has a low error tolerance) it may be preferable to use the set of interfaces that lead to long Good state bursts, as this results in the lower number of service interruptions. From the results obtained with a single interface, this would be LTE.

On the other hand, if an error tolerant service is considered, the set of interfaces that lead to the shortest bad state bursts may be preferable. From the results obtained with a single interface, this would be \mbox{Wi-Fi}.


The duration of Good state bursts relates directly to the concept of \emph{interval reliability} \cite{hossler2017applying,billinton1992reliability}, which is defined as the probability of uninterrupted service operation during a time interval of duration $\Delta t$. The longer the Good state bursts, the higher is the probability that the service is not interrupted during the interval $\Delta t$.

Considering now the diversity configurations, we see first that \emph{2x LTE} increases the Good state duration 10-fold (c.f. Fig.~\ref{fig:burst_length}(a)) relative to \emph{LTE} and halves the Bad state duration (c.f. Fig.~\ref{fig:burst_length}(b)). The \emph{2x \mbox{Wi-Fi}} excels with the shortest Bad state duration of all considered schemes and provides a surprisingly long Good state duration, only exceeded by \emph{2x LTE}.
Lastly, we see that the \emph{LTE + \mbox{Wi-Fi}} provides an interesting compromise, where the Good state duration is lower than \emph{2x LTE} but comparable to \emph{2x \mbox{Wi-Fi}}, and with a short Bad state duration that is much lower than LTE and close to that of \mbox{Wi-Fi}.

\subsection{A single reliable vs. two less reliable interfaces}
The second question we set out to investigate is how a single ultra-reliable interface compares to using interface diversity with two less reliable interfaces. To analyze this, we use the same methodology as above but estimate the Gilbert-Elliott model parameters differently, so as to imitate the normal and ultra-reliable cases. Specifically, we consider $p$ and $r$ determined for latency deadlines that lead to packet success rates of 0.95 (normal) and 0.995 (ultra-reliable), as given in Table~\ref{tab:est_p_r_ur}. 

\begin{table}[bt]
	\centering
	\caption{Estimated Gilbert-Elliott model parameters}
    \resizebox{\linewidth}{!}{
	\begin{tabular}{lccccccc}
	\toprule
	& \multicolumn{3}{c}{$0.95$} & & \multicolumn{3}{c}{$0.995$} \\
	& $l$ [ms] & $\hat{p}$ & $\hat{r}$ & & $l$ [ms] & $\hat{p}$ & $\hat{r}$ \\ \cmidrule{2-4} \cmidrule{6-8}
	LTE   & 38.7 & 0.0170252 & 0.374819 && 42.4 & 0.00450365 & 0.897892 \\
	Wi-Fi & 39.6 & 0.0499014 & 0.946788 && 92.9 & 0.00394534 & 0.774916  \\ \bottomrule
	\end{tabular}
	}
	\label{tab:est_p_r_ur}
\end{table}

\begin{figure}
    \centering
    \subfigure[Good state]{\includegraphics[width=\linewidth]{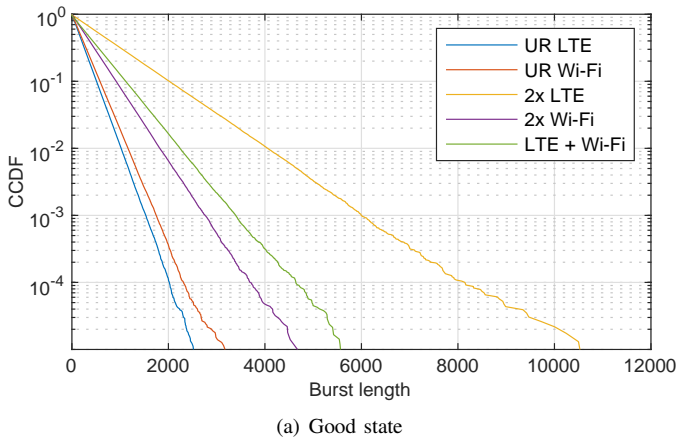}}
    \subfigure[Bad state]{\includegraphics[width=\linewidth]{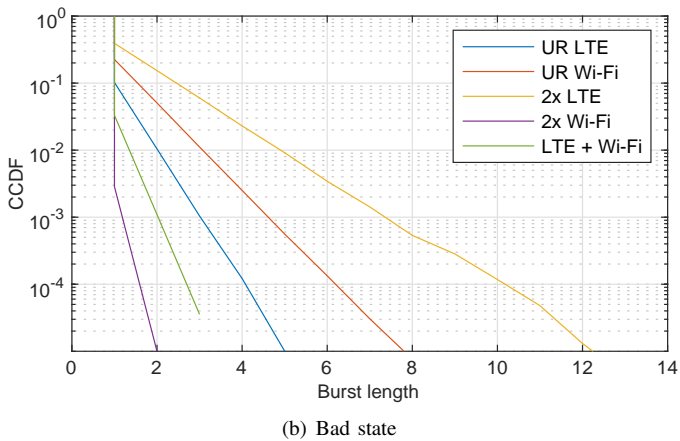}}
    \caption{Burst length distribution for Gilbert-Elliott model simulation with ultra-reliable (UR) single interfaces using parameters in Table \ref{tab:est_p_r_ur}.}
    \label{fig:burst_length_ur}
\end{figure}

The plot of Good state burst length distribution in Fig.~\ref{fig:burst_length_ur}(a) shows that the two ultra-reliable interfaces achieve the shortest Good burst lengths, as was the case in Fig.~\ref{fig:burst_length}(a). For the bad state length, the \emph{UR LTE} and \emph{UR \mbox{Wi-Fi}} are shorter than \emph{2x LTE}, though not as short as \emph{LTE + \mbox{Wi-Fi}} and \emph{2x \mbox{Wi-Fi}}.
In summary, our findings indicate that interface diversity configurations that include at least one \mbox{Wi-Fi} interface achieve the best performance in terms of Bad error burst length.

\balance
\section{Conclusions and Outlook}
\label{sec:conclusion} 
In this work we have presented an analysis of the impact of interface diversity on the  burst length of failed and successful packet transmissions. The analysis is based on simulations of a Gilbert-Elliott 2-state burst error model, where the transition probability parameters were estimated from latency measurements of LTE and \mbox{Wi-Fi}, subjected to different latency deadline values as a way to impose actual service requirements.

Our results show that interface diversity with packet duplication on two interfaces delivers 5-15 fold improvement in \emph{Good} state burst length, which relates to interval reliability, whereas the \emph{Bad} state (error) burst length of LTE can be reduced by 90\% by pairing LTE with \mbox{Wi-Fi}.
Secondly, we also compared interface diversity configurations to ultra-reliable single interfaces. We found that two-interface diversity configurations including \mbox{Wi-Fi} were always superior to ultra-reliable LTE or \mbox{Wi-Fi} in terms of burst length statistics, whereas LTE dual connectivity was only superior, but by a large margin, in \emph{Good} state burst length (interval reliability).

In conclusion we emphasize that the study is based on latency measurement traces obtained in a specific location with specific environmental conditions. Hence, our findings are not necessarily generic and should be seen as indicative---encouraging further studies in this direction. 

\section*{Acknowledgment}
The work was supported in part by the European Research Council (ERC Consolidator Grant no. 648382 ''WILLOW'') and by the European Union's research and innovation programme under the Marie Sklodowska-Curie grant agreement No. 765224 ''PAINLESS'' within the Horizon 2020 Program.

\bibliographystyle{IEEEtran}

\end{document}